\pdfoutput = 1
\documentclass[
5p
]{elsarticle}

\usepackage{hyperref}
\usepackage{amsmath}

\usepackage{tabu}
\usepackage{booktabs}
\usepackage{geometry}
\usepackage{caption}
\usepackage{subcaption}
\usepackage{siunitx}

\journal{Journal of Vascular Surgery}

\biboptions{super,sort&compress}

\begin{document}

\begin{frontmatter}

\title{Partial renal coverage in EVAR causes unfavorable renal flow patterns in an infrarenal aneurysm model}

\author[affiliation1,affiliation2]{L. van de Velde}
\author[affiliation1]{E.J. Donselaar}
\author[affiliation1,affiliation2,affiliation3]{E. Groot Jebbink}
\author[affiliation1,affiliation4]{J.T. Boersen}
\author[affiliation2,affiliation3]{G.P.J. Lajoinie}
\author[affiliation4]{J.P.M. de Vries}
\author[affiliation5]{C.J. Zeebregts}
\author[affiliation2,affiliation3]{M. Versluis}
\author[affiliation1]{M.M.P.J. Reijnen
\corref{mycorrespondingauthor}}
\cortext[mycorrespondingauthor]{Corresponding author at:
Department of Surgery, Rijnstate Hospital, Wagnerlaan 55, 6815 AD Arnhem, The Netherlands.}
\ead{mmpj.reijnen@gmail.com}

\address[affiliation1]{Department of Surgery, Rijnstate Hospital, Arnhem, The Netherlands}
\address[affiliation2]{MIRA Institute for Biomedical Technology and Technical Medicine, University of Twente, Enschede, The Netherlands}
\address[affiliation3]{Physics of Fluids group, Faculty of Science and Technology, University of Twente, Enschede, The Netherlands}
\address[affiliation4]{Department of Vascular Surgery, St. Antonius Hospital, Nieuwegein, The Netherlands}
\address[affiliation5]{Department of Surgery, Division of Vascular Surgery, University Medical Center Groningen, University of Groningen, Groningen, The Netherlands}

\begin{abstract}
\textbf{Objective} To achieve an optimal sealing zone during EVAR, the intended positioning of the proximal end of the endograft fabric should be as close as possible to the most caudal edge of the renal arteries. Some endografts exhibit a small offset between the radiopaque markers and the proximal fabric edge. Unintended partial renal artery coverage may thus occur. This study investigates the consequences of partial coverage on renal flow patterns and wall shear stress. 

\noindent \textbf{Methods} In-vitro models of an abdominal aortic aneurysm were used to visualize pulsatile flow using 2D particle image velocimetry under physiologic resting conditions. One model served as control and two models were stented with an Endurant endograft, one without and one with partial renal artery coverage with 1.3 mm of stent fabric extending beyond the marker (16\% area coverage). The magnitude and oscillation of wall shear stress, relative residence time and backflow in the renal artery were analyzed.

\noindent \textbf{Results} In both stented models, a region along the caudal renal artery wall presented with low and oscillating wall shear stress, not present in the control model. A region with very low wall shear stress (\SI{<0.1}{Pa}) was present in the model with partial coverage over a length of 7 mm, compared to a length of 2 mm in the model without renal coverage. Average renal backflow area percentage in the renal artery incrementally increased from control (0.9\%) to the stented model without (6.4\%) and with renal coverage (18.8\%).

\noindent \textbf{Conclusion} In this flow model partial renal coverage after EVAR causes low and marked oscillations in wall shear stress, potentially promoting atherosclerosis and subsequent renal artery stenosis. Awareness of the device-dependent offset between the fabric edge and the radiopaque markers is therefore important in endovascular practice. (J Vasc Surg 2018;67:1585–1594)

\noindent \textbf{Clinical Relevance} The location of the proximal markers of EVAR devices is not always at the most proximal site of the graft material and as such part of the coverage material may be placed over the orifice of the renal artery. A recent analysis showed an incidence of inadvertent partial coverage of 28\%. The association of a slight renal artery coverage with adverse flow patterns in this study stresses the importance of avoiding any renal artery coverage during deployment of the main body. The results further act to emphasize the importance of optimal C-arm adjustment to obtain a perpendicular projection of the lowermost renal artery.
\end{abstract}

\end{frontmatter}

\section{Introduction}
\begin{figure}
\centering
	\subcaptionbox{\label{fig:1A}}
		[.49\linewidth]{\includegraphics[width=.485\linewidth]{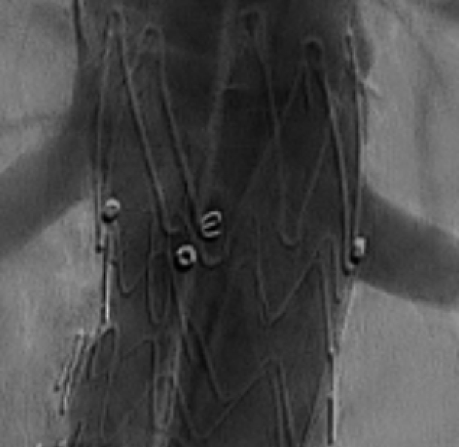}}
	\subcaptionbox{\label{fig:1B}}
		[.49\linewidth]{\includegraphics[width=.48\linewidth]{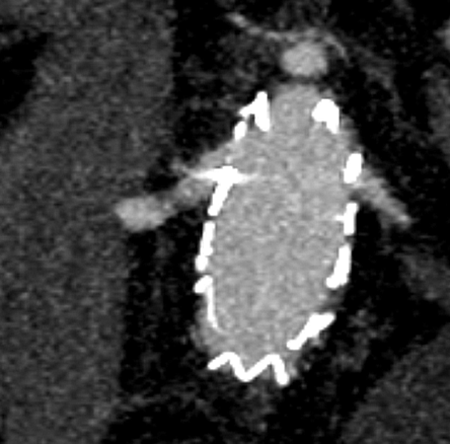}}
\caption{\label{fig:1}Two cases of partial renal coverage in our hospital after EVAR with Medtronic Endurant stent grafts. A, Angiography after EVAR with partial coverage of the renal artery, with the marker positioned clearly above the caudal border of the renal artery orifice. B, Coronal CTA reconstruction after EVAR with the high-intensity marker positioned immediately below the renal artery orifice. With the Medtronic Endurant stent, partial renal artery coverage by 1-1.5 mm is present in this case.}
\end{figure}

Advantages of endovascular aneurysm repair (EVAR) over conventional open surgical abdominal aortic aneurysm (AAA) repair include a lower 30-day mortality rate (1.7\% vs. 4.2\%) and a shorter recovery time (3-10 vs. 7-16 days). A higher re-interventation rate is the most important drawback of EVAR.\citep{Paravastu2014} Although EVAR is associated with a lower risk of acute renal failure compared to open surgery (adjusted odds ratio 0.42, 95\% confidence interval [0.33 to 0.53] ),\citep{Wald2006} both treatments have similar renal complication rates in the mid-term.\citep{Paravastu2014} Acute kidney injury or hemodialysis requirement occurred in 3.3\% patients after EVAR in a retrospective analysis,\citep{Zarkowsky2016} associated with a more than 20\% decrease in 5-year survival, also when adjusted for differences in pre-operative eGFR. 
The mechanisms by which EVAR influences renal function are probably multifactorial. Thromboembolization, MRI and CT contrast media usage, stent fixation\citep{Sun2006,Miller2015} and ischemic reperfusion injury are potential factors.\citep{Walsh2008a} In addition, unintended partial coverage of a renal ostium with covered stent material may occur,\citep{VanDijk2003} both as a result of an unintended overly cranial release of the endograft and also as a result of inaccurate positioning of the radiopaque marker on the endograft. In many endografts the covered stent material extends about 1 mm beyond their proximal radiopaque markers and positioning the marker flush below the renal artery will lead to partial coverage by the graft material above the marker (Fig. \ref{fig:1}). The position of the radiopaque marker is not mentioned by some manufacturers in their instructions for use (IFU) and the exact distance to covered material may therefore be unknown to interventionalists. The extent to which the renal orifice is blocked varies in a range of 0–2 mm for the different endografts (Table I) and may not always be appreciated on completion angiography.\citep{VanDijk2003} In a recent postoperative CT analysis of EVAR procedures even 28\% of all endograft placements were reported to partially cover the renal artery.\citep{Schuurmann2017}
Partial coverage of the renal orifice may induce unfavorable hemodynamics such as regions of low and oscillatory flow. Low levels of wall shear stress (WSS \SI{<<1}{Pa}), in addition to alternating flow directions throughout the cardiac cycle quantified by an oscillatory shear index (OSI),\citep{Ku1985} have been associated with early atherosclerosis.\citep{Moore1994b,He1996,Gimbrone2013a} Based on atherosclerotic pathophysiology, WSS and OSI can be combined into a single risk factor, the relative residence time (RRT).\citep{Himburg2004,Lee2009} Furthermore, flow separation and recirculating flow with an associated region of reversed flow have been suggested as pro-atherogenic\citep{Gimbrone2013a,Bluestein1996} and can impede inflow at arterial branches.\citep{Katritsis2010} The present in-vitro study investigates the effect of partial renal artery coverage by graft material on renal flow patterns and WSS in the renal artery.

\begin{table}
\footnotesize   
\caption{\label{tab:1}Distance between proximal marker center and edge of stentgraft}
\centering
\sisetup{range-phrase = --}
\begin{tabu} to\linewidth{@{} X[1] X[-1] X[2 c] @{}} 
\toprule
Device 		& Manufacturer 					& Marker distance (mm)\\ 
\midrule 
Endurant	& Medtronic   						&	\numrange{1}{1.5}		\\
Talent	 	& Medtronic        					& 	0 				 				\\
Excluder   	& Gore 									&	\numrange{1.2}{1.7}  	\\
Zenith       	& Cook 									& \num{<2}	 							\\
Anaconda	& Vascutek 							& 0 								\\
\bottomrule
\end{tabu}
\caption*{Offset of proximal radiopaque marker in common endovascular aneurysm reapir (EVAR) endografts as described in the manufacturer's instruction for use (Medtronic, Cook) or, for those not mentioned in the instructions for use, by our measurements (Gore, Vascutek)}
\end{table}

\section{Methods}
\subsection{In-vitro flow models}
\begin{figure}[h]
\centering
\includegraphics[width=\linewidth]{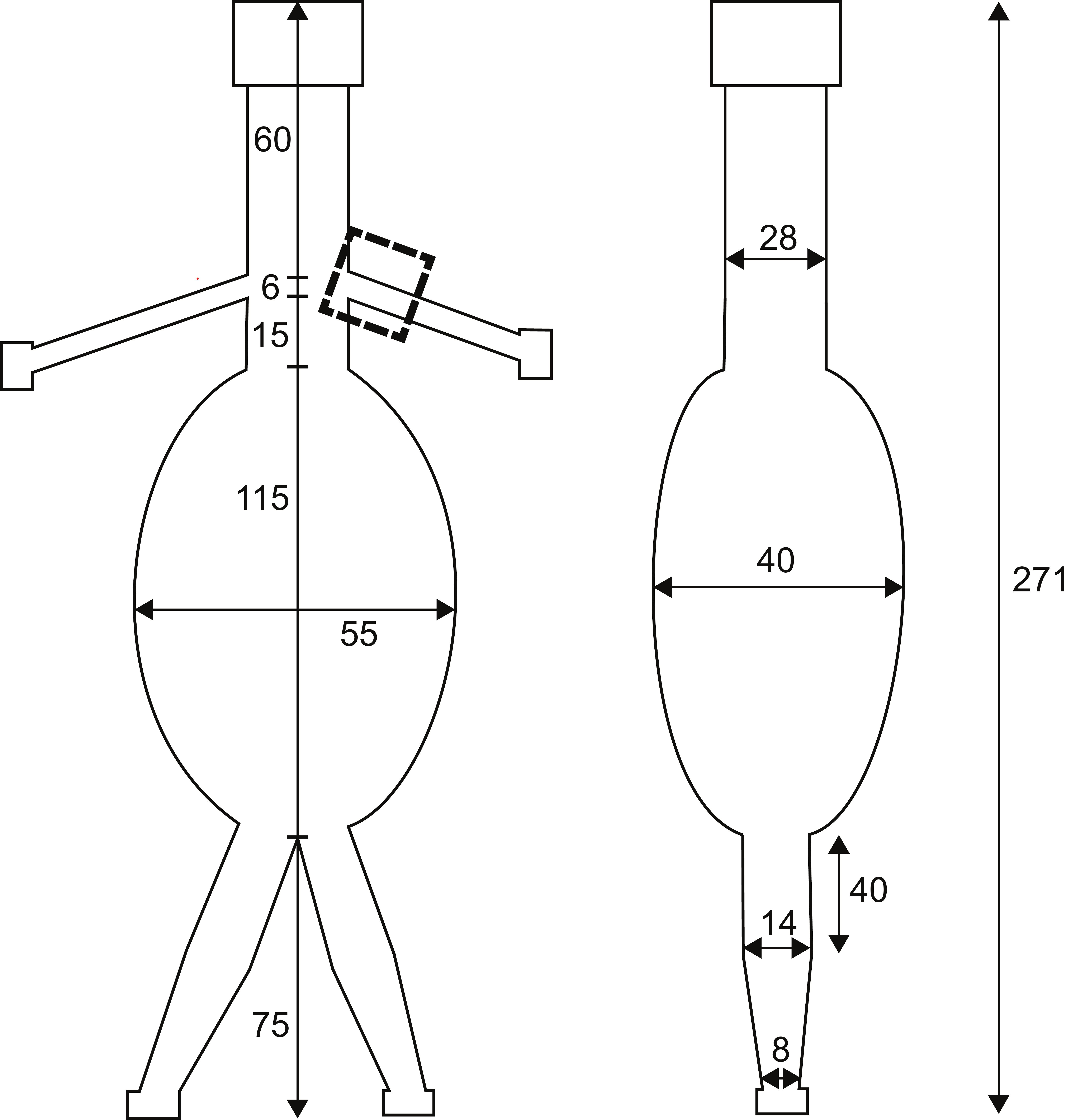}
\caption{\label{fig:2}Anatomic parameters of the AAA model in a frontal (left) and sagittal (right) plane. The square indicates the field of view of the PIV camera on the renal artery. The shown sizes are in mm. Take-off angles are \ang{60} for both renal arteries, \ang{30} for the right iliac artery and \ang{20} for the left iliac artery.}
\end{figure}

Three transparent silicone flow phantoms were used as in-vitro models of the infrarenal abdominal aorta. The manufacturing process of these phantoms has been previously described.\citep{GrootJebbink2015} Characteristics of the anatomy are shown in Fig. \ref{fig:2}. An aneurysm size of \SI{55}{mm} in the lateral direction was set to represent an AAA for which elective treatment is indicated. The anteroposterior diameter was restricted by our manufacturing equipment and was set to \SI{40}{mm}. Anteroposterior angulation was not considered when designing the model to facilitate 2D-imaging of the flow in the midline plane. In addition, the inferior mesenteric artery was not included and orifices of left and right renal arteries were situated at equal height. 
One control model and two configurations with an implanted Endurant II Stent Graft System (Medtronic Inc., Minneapolis, MN, USA) were studied. The endografts were implanted in the model by an experienced vascular surgeon (MR). For the first configuration, the radiopaque marker at the proximal site was positioned \SI{1}{mm} below the renal orifices following the manufacturers' IFU. For the second configuration, the radiopaque marker was positioned directly at the inferior border of the renal orifices, causing partial coverage of the renal ostia with 1.3 mm of covered stent material that extended beyond the radiopaque marker. Due to the \ang{60} take-off angle of the renal artery, the renal ostium is an ellipse with a short axis of \SI{6}{mm} (equal to the renal artery diameter) and a long axis of \SI{6.9}{mm} (6/cos(\ang{30})). The graft material covers \SI{1.3}{mm} of the short axis, corresponding to a 16\% area coverage of the renal ostium. 

\subsection{Flow circuit}
To simulate physiologic flow levels a previously described flow setup was used.\citep{GrootJebbink2014} Flow was driven by three gear pumps generating a pulsatile flow. Before delivery to the inlet of the model, the flow traversed through a tube sufficiently long for a fully developed pulsatile flow profile to develop. A two-element Windkessel consisting of a resistive and capacitive component was used to model the vascular bed distal to the iliac arteries. The load on the renal arteries was modeled with a resistive element. Studies were performed under simulated resting conditions at a heart rate of 60 beats per minute and a mean suprarenal aortic flow rate of \SI{1.6}{L.min^{-1}} with equal time-averaged volume flow of \SI{400}{mL.min^{-1}} in each of the renal and iliac arteries. These values correspond to physiologic resting conditions used in previous studies.\citep{Moore1994} Overal resistance was set for mean distal iliac pressure to equal 100 mmHg. The capacitive element distal to the iliac arteries was set such that the distal iliac pressure ranged between 80 and 120 mmHg. A blood mimicking fluid (BMF) with a dynamic viscosity of 4.3 mPa.s and a refractive index of 1.4, matching the refractive index of the silicone flow phantom, was produced by mixing water, glycerol and sodium iodide in a weight-ratio of 47.4\%, 36.9\% and 15.7\%.\citep{Yousif2011} Fluorescent Poly Methyl Methacrylate (PMMA) particles (Rhodamine, size 1-20 \si{\mu m}, density \SI{1190}{kg.m^{-3}}, Dantec Dynamics A/S, Skovlunde, Denmark) were added to the BMF for flow quantification.

\subsection{Flow quantification}
For visualization, the output of a continuous wave laser (5W DPSS laser, 532 nm, Cohlibri, Lightline, Germany) was focused to form a laser sheet with a width of 1 mm to illuminate a thin layer of liquid. The laser sheet was aligned with the center of the flow lumen of the right renal artery in the frontal plane. A high-speed camera (FASTCAM SA-X2, Photron Inc., West Wycombe, Buckinghamshire UK) was positioned perpendicular to the illuminated in-plane layer and equipped with an optical notch filter (monochromatic 532 nm, Edmund optics, Barrington, NJ, USA) to capture the fluorescent signal emitted by the microbeads at a frame rate of 2000 frames per seconds. The distance between the camera and the imaged plane was close to the focal length of the optical setup to allow accurate PIV measurement of high velocities. With a field of view of 28x28 mm, a frame size of 1024x1024 pixels, and an interrogation window of 64x64 pixels, velocities up to 1.75 m/s could be correctly captured.
The images were background subtracted and pre-processed with contrast-limited adaptive histogram equalization and a high-pass intensity filter. Particle image velocimetry (PIV) with an iterative multigrid approach\citep{Scarano1999} was perfomed using the Matlab (Mathworks Inc., MATLAB R2016a, Nattick, MA) toolbox PIVlab\citep{Thielicke2014} (version 1.41). After the PIV analysis, data outside of absolute, relative and local median based velocity tresholds were discarded. Missing values were interpolated from neighboring values using a boundary value solver.\citep{Thielicke2014} Finally, velocity data was averaged over 10 cardiac cycle to produce average velocity vectors throughout one cardiac cycle. With a low measurement variability for the 10 cycles, of similar as demonstrated in one our previous studies,\citep{Boersen2017b} the pulsatile inflow profile was accurately reproduced for all three models.

\subsection{WSS and backflow quantification}
The shear rate at both walls of the renal artery was computed with MATLAB in-house built scripts. First, a parametric expression (first-order polynomial) of these walls was created with contrast-based edge detection of the border between regions of flow (the vessel lumen) and no flow. Then, at points in the wall closest to a velocity vector, a normal line to the wall was constructed, resulting in a spacing between normal vectors of about 16 pixels (0.44 mm). The velocity vector field from the PIV analysis was interpolated with a natural neighbor scheme (Matlab's ScatteredInterpolant) for points on the normal vectors. For every normal wall vector, a cubic spline was fitted. Velocity magnitude at the wall was constrained at zero, i.e. a no-slip boundary condition. Shear rate was computed as the derivative of the fluid velocity component parallel to the wall, in the spatial direction normal to the wall. 

\begin{table*}
\footnotesize   
\caption{\label{tab:2}Calculation of hemodynamic parameters that are associated with atherosclerosis}
\centering
\begin{tabu} to\linewidth{@{} X[-1] X[2] X[1 $c ] @{}} 
\toprule
		 		& Description 					& \textrm{Equation}\\ 
\midrule 
$\tau_w$		& WSS is the product of dynamic viscosity times the derivative of the parallel wall velocity in the direction perpendicular to the wall   						&	\mu \frac{\partial v_i}{\partial x_j}		\\
TAWSS	 	& Average WSS magnitude over one heart cycle        					& 	\frac{1}{T}\int_0^T \left| \tau_w \right| \textrm{d} t 				 				\\ 
\addlinespace[1mm]
OSI   		& Index WSS alternation between negative and positive shear (0 for no oscillation up to 0.5 for maximal oscillation) 									&	\frac{1}{2} \left( 1- \frac{\int_0^T \left| \tau_w \right| \textrm{d} t }{ \left|\int_0^T \tau_w  \textrm{d} t \right|} \right)  	\\
RRT       	& Theoretical relative residence time of blood solutes and particles near the wall 									& \frac{1}{1-2 \textrm{OSI}) \left| \textrm{TAWSS} \right|}	 							\\
\bottomrule
\end{tabu}
\caption*{OSI, Oscillatory shear index; T, time of one heart cycle, $\tau_w$, instantaneous wall shear stress; TAWSS, time-averaged wall shear stress; RRT, relative residence time; WSS, wall shear stress.}
\end{table*}

\begin{figure}[h]
\centering
\includegraphics[width=\linewidth]{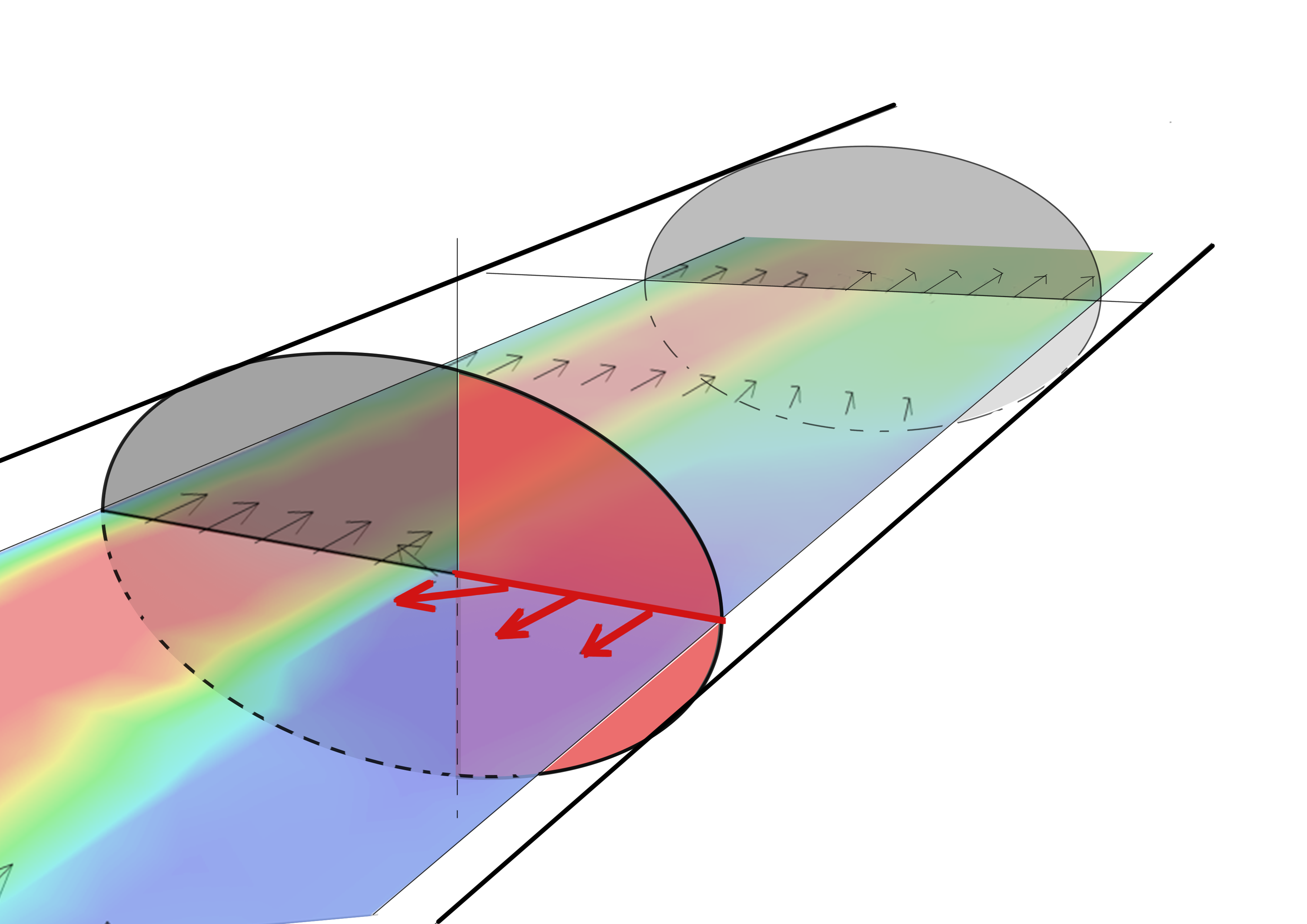}
\caption{\label{fig:3}Illustration of the renal backflow area (RBA) calculation. Flow has been visualized in a frontal cut-plane of the cylindrical renal artery. For an axial cross-section where backflow is present, the cross-sectional area with backflow, shown as red area, is calculated. RBA equals the ratio of this backflow area to the full circular area of the cross-section.}
\end{figure}

Time-averaged wall shear stress (TAWSS), OSI\citep{He1996} and RRT\citep{Himburg2004} were calculated along the length of the renal walls, according to the equations shown in Table II. The extent of backflow in the renal artery was quantified analogous to a study in a coronary bifurcation\citep{Katritsis2010} by determining the percentage of area having velocity vectors facing opposite to the forward flow direction. First, the normal line with negative velocities extending furthest from the wall per timepoint was selected. This provided a radial length along which backflow is present. With the assumption that in 3D backflow was present at sections above and below the laser sheet in the renal artery cross-section, the part of the circular cross-section where backflow was present was calculated. Fig. \ref{fig:3} show the two areas used for this calculation. The resulting percentage was termed renal backflow area (RBA).

\section{Results}
\begin{figure*}
\centering
	\subcaptionbox{\label{fig:4A}}
		[.24\linewidth]{\includegraphics[width=.235\linewidth]{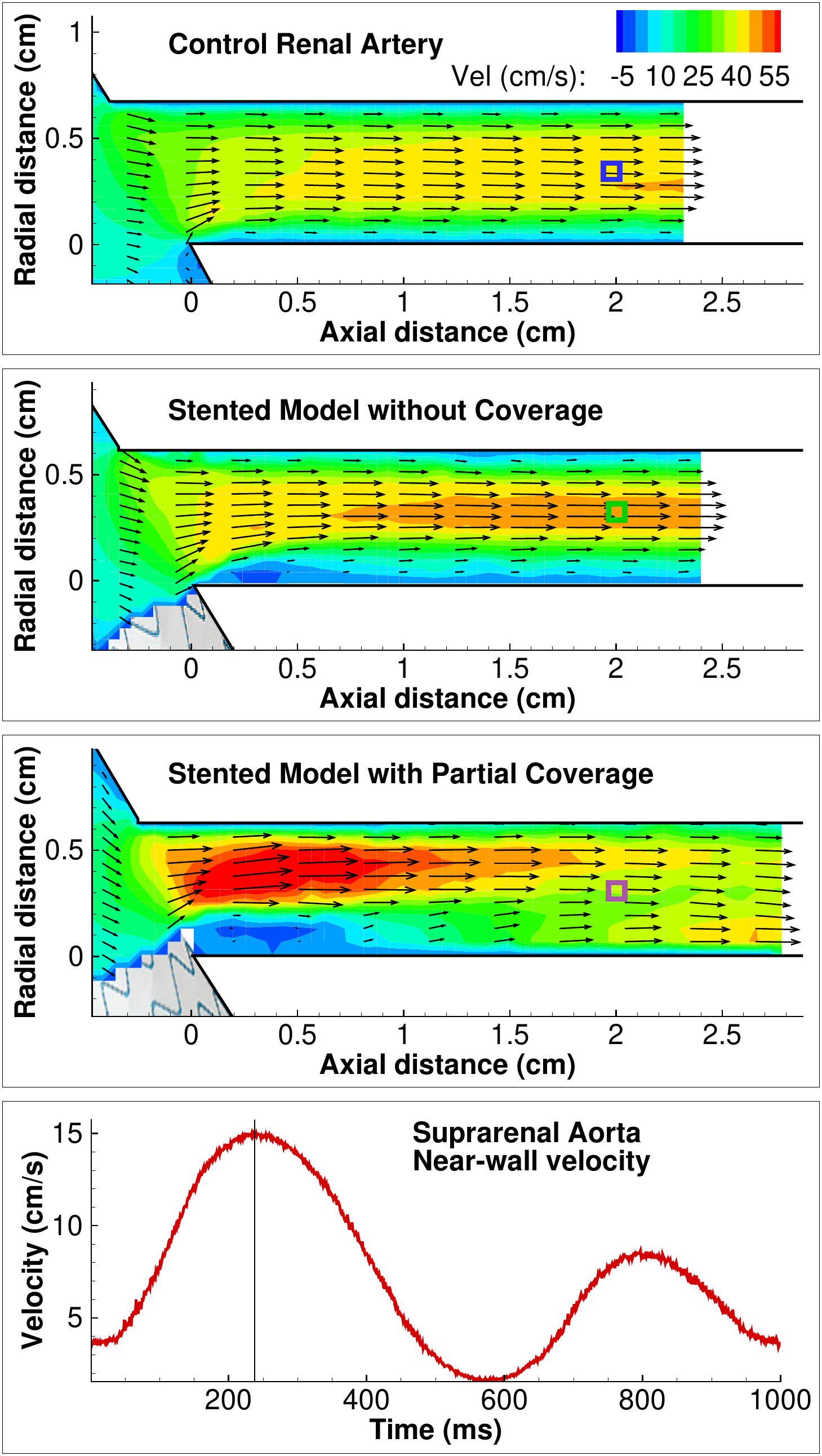}}
	\subcaptionbox{\label{fig:4B}}
		[.24\linewidth]{\includegraphics[width=.235\linewidth]{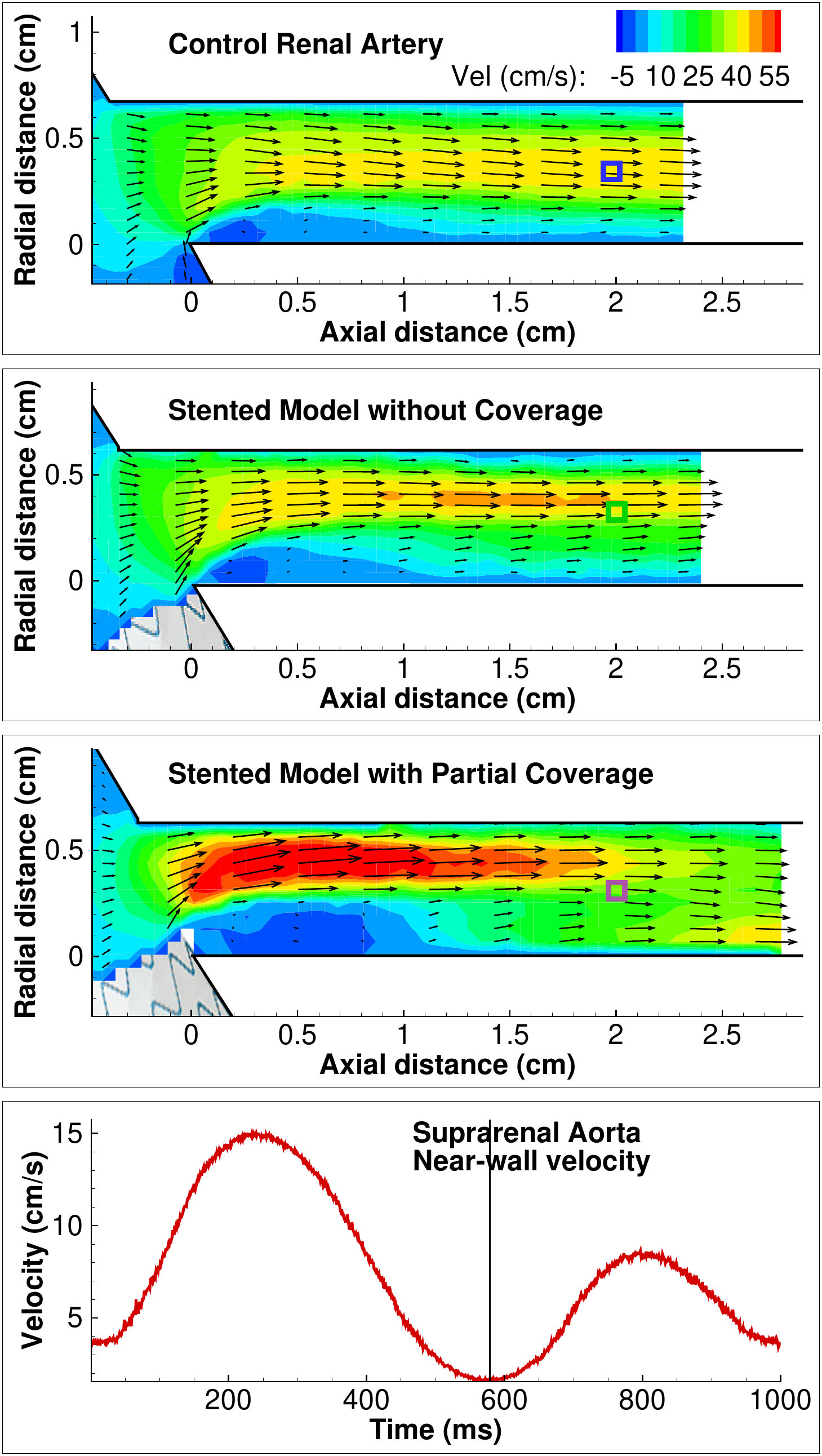}}
	\subcaptionbox{\label{fig:4C}}
		[.24\linewidth]{\includegraphics[width=.235\linewidth]{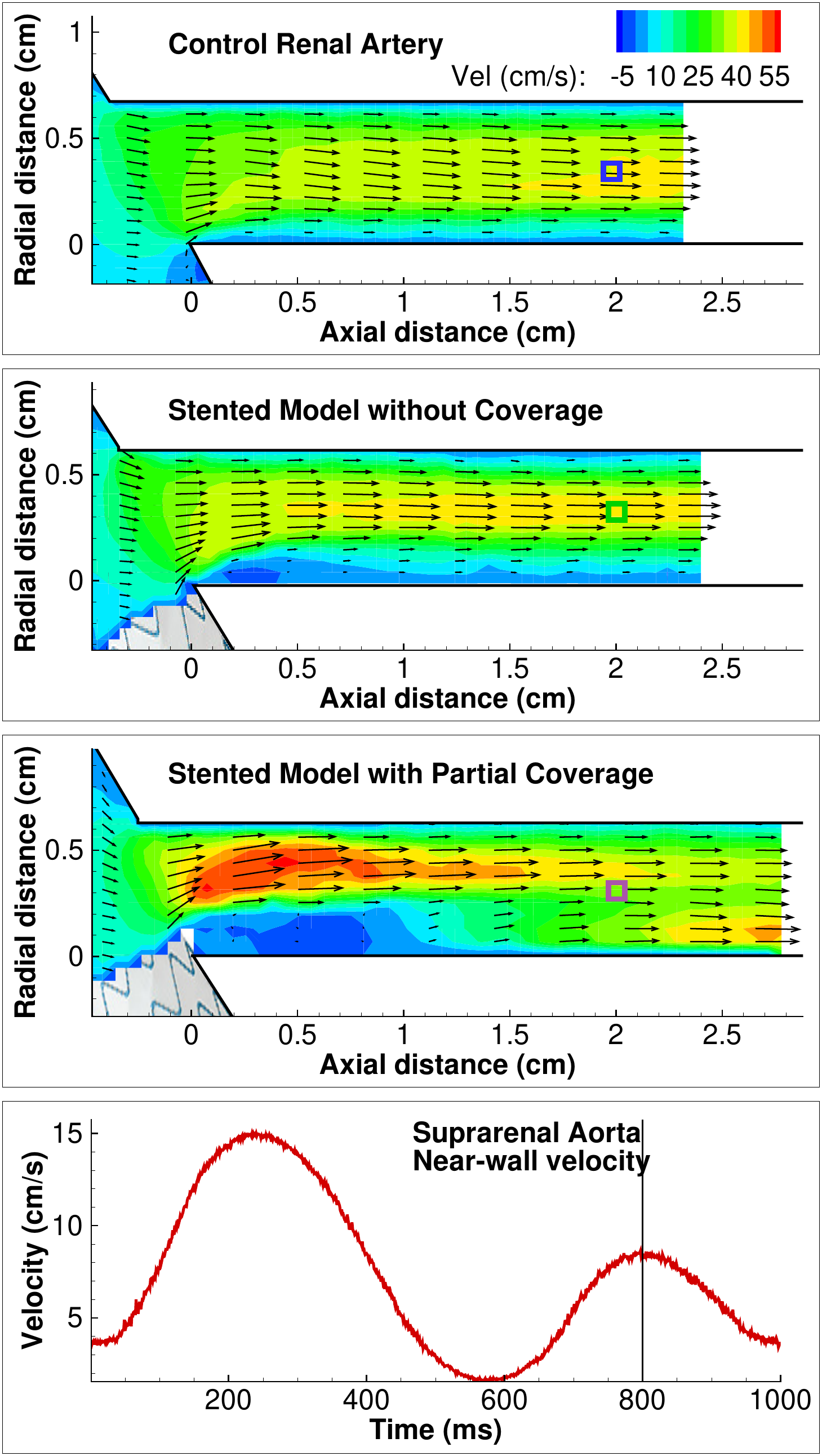}}
	\subcaptionbox{\label{fig:4D}}
		[.24\linewidth]{\includegraphics[width=.235\linewidth]{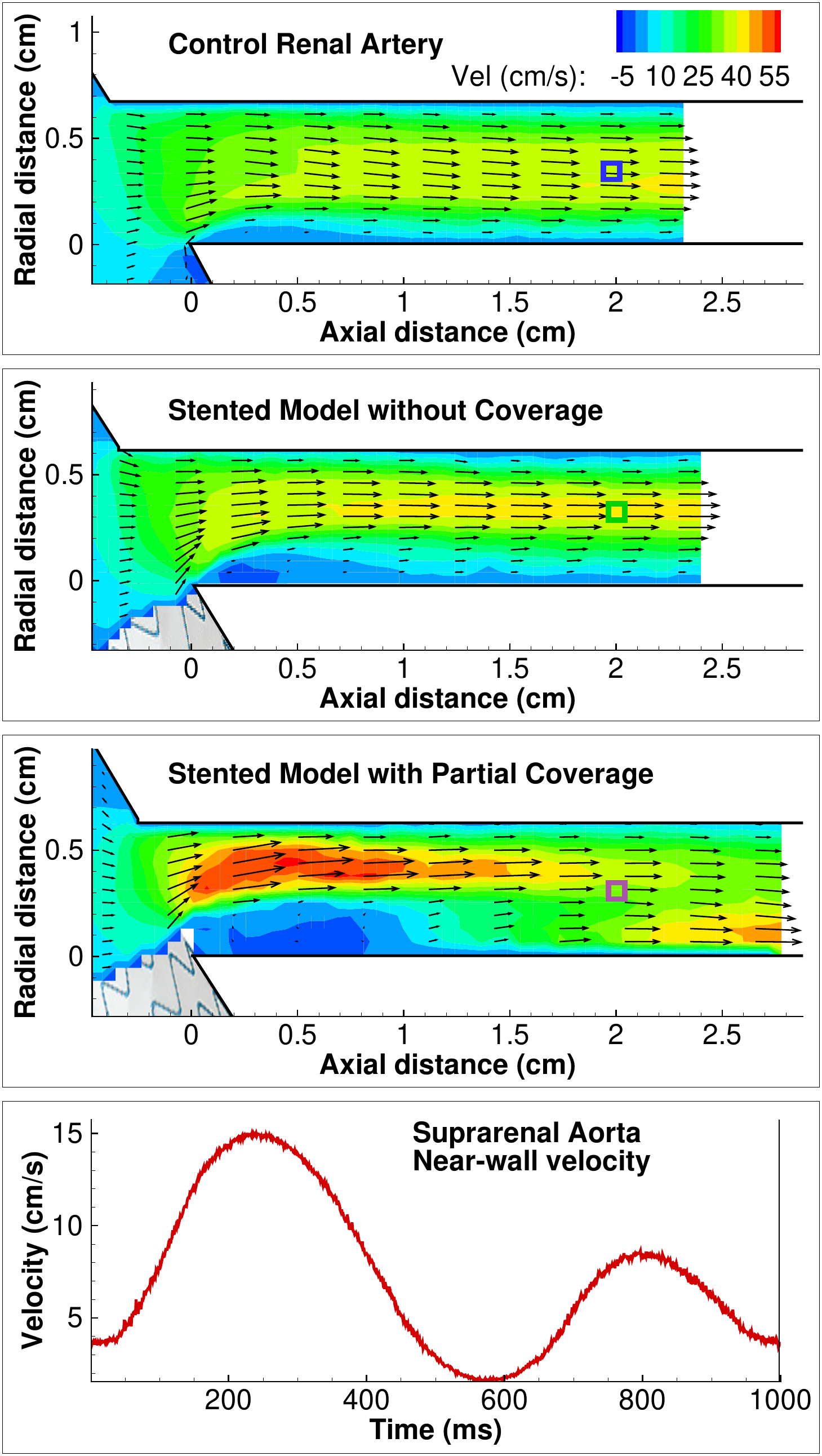}}
\caption{\label{fig:4}Flow patterns in the three models at the onset of the renal artery. A small part of the juxtarenal aorta is visible on the left, with the branching renal artery aligned with the horizontal axis. Approximate position of the infrarenal endograft is shown. The blue squares at axial distance 2 cm indicate the position at which time-dependent central lumen velocity is plotted in Fig. \ref{fig:5}. The bottom graph shows near-wall instantaneous velocity at the suprarenal aortic wall to show the systolic and diastolic phases, but is not indicative of aortic volumetric flow. (a) Peak flow systole, (b) End systole, (c) Peak flow diastole (d) End diastole.}
\end{figure*}

\begin{figure}[h]
\centering
\includegraphics[width=\linewidth]{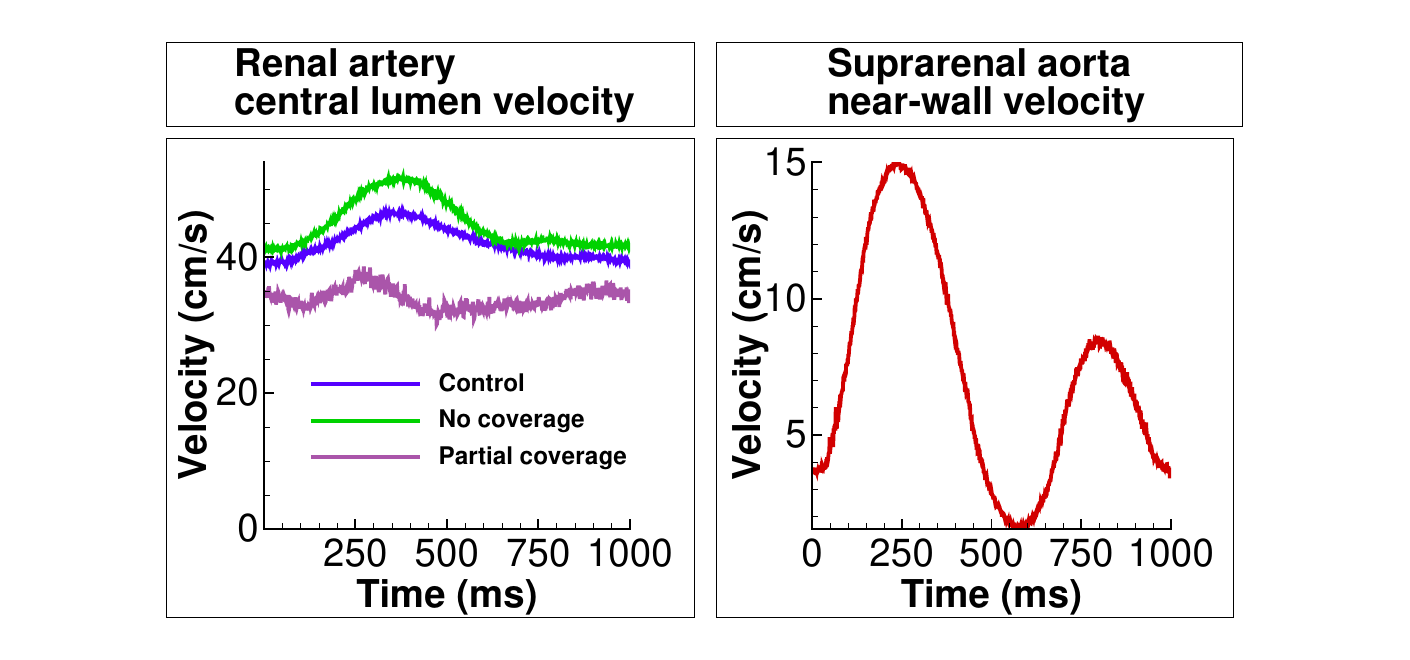}
\caption{\label{fig:5}Left: central lumen velocity over time in the three models at an axial distance 2 cm after the onset of the renal artery, as marked by the colored squares in Fig. \ref{fig:4}. Right: near-wall velocity in the suprarenal aorta showing the systolic and diastolic phases.}
\end{figure}

\sisetup{range-phrase = --}
The axial velocity contours in Fig. \ref{fig:4} provide an {over-view} of the flow during peak systole, end systole and diastole in the renal artery. Based on inlet centerline velocity and renal artery diameter, Reynolds numbers (mean-max inlet velocity) were \numrange{600}{690}, \numrange{670}{740} and \numrange{1000}{1100} in the control, stented model without and stented model with partial renal artery coverage. The flow profiles for the complete cardiac cycle are available in the online supplementary video S1. Instantaneous central lumen velocity in the renal artery was relatively constant in time when compared to the pulsatile suprarenal flow variations, as shown in Fig. \ref{fig:5}. In the stented model with partial renal artery coverage, central lumen velocity shows high-frequency oscillations not present in the other two models.  
The control and stented model without partial coverage displayed flow profiles with peak velocity close to the center, with maximal velocities of \num{47} and \SI{52}{cm.s^{-1}} halfway the deceleration phase of systole, respectively. The stented model with partial coverage was characterized by the formation of a jet with peak velocity of \SI{71}{cm.s^{-1}} near the cranial wall. Further downstream the jet was no longer visible in the imaged plane and a peak velocity near the caudal wall was visible during the cycle. The flow patterns remained qualitatively similar during the cardiac cycle, except for the flow separation, i.e. reverse flow near the wall, near the entrance. For the control model, flow separation was present only during and shortly after the deceleration phases of inlet flow. For both stented models, the flow separation was present during the complete cycle and the resulting region of stagnant or recirculating flow was maximal during deceleration. The region with reversed flow was consistently longer in the model with partial coverage compared to the model without partial coverage (about \SI{9}{mm} vs. \SI{3}{mm} at end systole). 

\begin{figure*}[h]
\centering
\includegraphics[width=\linewidth]{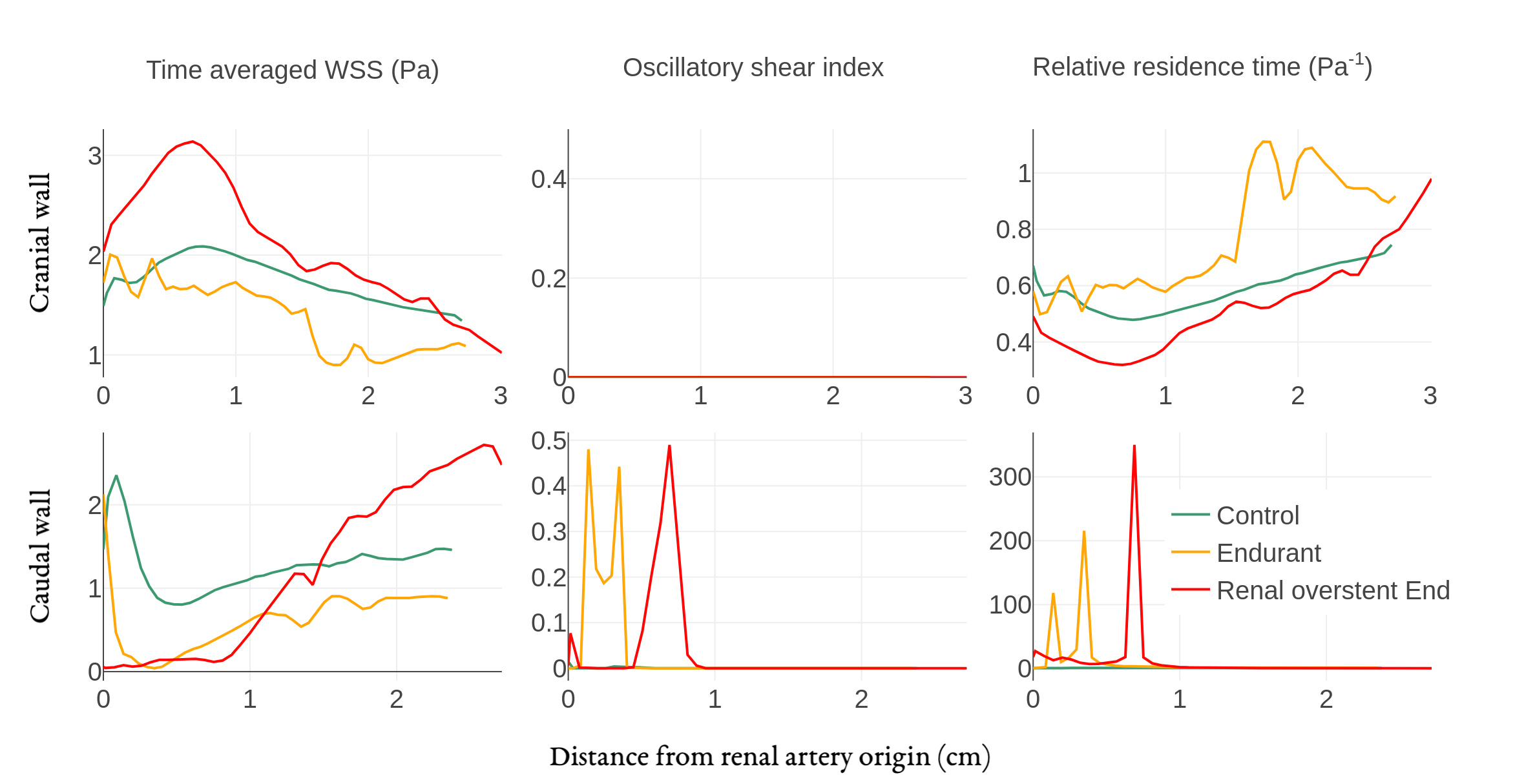}
\caption{\label{fig:6}Indices of WSS along the caudal and cranial renal wall. OSI is zero along the cranial wall for all models, as no negative flow was present.}
\end{figure*}

Shear rates in the models peaked at \SI{732}{s^{-1}} (control), \SI{820}{s^{-1}} (model without partial coverage) and \SI{1670}{s^{-1}} ({mo-del} with partial coverage). Indices of WSS along the renal artery walls for a full heart cycle are displayed in Fig. \ref{fig:6}. For the cranial wall, TAWSS was in the range of \numrange{1}{3} \si{Pa} for all models. Backflow was not present near the cranial walls in any of the models, so oscillatory shear was uniformly zero and RRT depicts the inverse of TAWSS. Near the origin of the renal artery, TAWSS was relatively high in the model with partial coverage, corresponding to high flow velocities observed in the flow patterns. For the caudal wall, negative and oscillating WSS was present in both stented models as opposed to control. For the model without partial coverage, TAWSS is low (\SI{<0.4}{Pa}) over a wall length of \SI{7}{mm} with high OSI values at the first half of this region. For the model with partial coverage, TAWSS was close to zero over a length of 7 mm starting from the orifice, with an additional region of 3 mm with low TAWSS (\SI{<0.4}{Pa}). RRT was large at these points on the caudal wall with a peak value of \SI{1700}{Pa^{-1}} for the stented model with partial coverage.
Quantifying backflow in the renal arteries as RBA gave a time-averaged RBA over a cardiac cycle of 0.9\% (max 12.6\%) for control, 6.4\% (max 13.1\%) for model without partial coverage and 18.8\% (max 31.3\%) for the model with partial coverage. Time incidences of maximum RBA were in the transition phase from systole to diastole for all models.

\section{Discussion}
With an in vitro analysis of flow patterns in the proximal renal artery, we have shown that partial renal artery coverage during EVAR procedures does affect local flow patterns. With \numrange{0.5}{1} \si{mm} graft coverage of a 6-mm renal artery, a 1-cm region along the caudal renal wall with low and oscillating WSS was observed and on average 18.8\% of the cross-sectional area exhibited flow reversal with a maximum of over 30\% at the end of systole. 
Although the pathophysiological mechanisms of WSS in vascular pathology are still not fully elucidated, there is consensus that areas of low and oscillating WSS contribute to the initiation and development of atherosclerosis.\citep{Peiffer2013a} In line with recent studies in the field,\citep{Lee2009,Morbiducci2010,Suess2016} we have used RRT as a measure to account for both these effects. In both stented models a small part along the caudal renal wall displayed extremely elevated RRT (\SI{>2000}{Pa^{-1}} for the model without partial coverage and \SI{>17000}{Pa^{-1}} for the model with partial coverage). These were mainly caused by a high oscillatory shear at these sites, i.e. the contribution of negative flow roughly equals the contribution of positive flow over a single heart cycle. Theoretically, the low velocity associated with low WSS along with the high oscillatory degree promotes an oscillating movement of blood particles over a small area of the wall, potentially inducing near-wall interactions.\cite{Himburg2004} The substantial elevation in RRT associated with renal coverage emphasizes the importance of meticulous positioning of the endograft, for which accuracte C-arm positioning perpendicular to the lowermost renal artery take-off is essential. 
Opposed to the long-term WSS related effects, high shear rates and renal flow impairment could cause clinical complications on a short timescale. Shear rates exceeding \SI{5000}{s^{-1}} are associated with acute arterial thrombo-embolization\citep{Casa2015} which could be an additional explanation for the slight increase in renal infarction seen with EVAR compared to conventional surgery.\citep{Sun2006} However, in the current models the shear rates with a maximum of \SI{1670}{s^{-1}} are far below the critical threshold. In EVAR procedures with a higher degree of renal coverage than the current study, the associated increase in WSS could contribute to plaque development or rupture in pre-existent plaques in the proximal part of the cranial side of the renal artery.\citep{Eshtehardi2017} However, the mechanical forces exerted on the aortic wall during endograft introduction could induce thrombo-embolization and may be a more frequent cause of renal infarction. 
When considering backflow as a region of functional flow impairment, a stent with and without renal coverage respectively block the renal cross-sectional flow area by 6.4\% and 18.8\% respectively. Especially during diastole, partial renal artery coverage causes backflow in 31.3\% of the renal lumen as opposed to 12.6\% in control conditions. A potential mechanism by which the stents can increase renal backflow is by reducing diastolic backflow from the infrarenal aorta into the renal arteries. Normally, the steep pressure drop at the end of systole leads to flow separation and backflow at the infrarenal wall, freely moving into the low resistance renal arteries.\citep{Holenstein1988} With an endograft in situ, the pressure drop along the infrarenal wall is increased, potentially decreasing the adverse pressure gradient during diastole and thus limiting flow separation and backflow into the renal arteries. Flow visualization in an aorta with a Cordis AAA endograft deployed confirmed that the endograft augments flow disturbances during flow deceleration and decreases inflow to the renal arteries at rest.\citep{Walsh2003} In addition, the endograft bifurcates the flow at a more cranial position, which may alter the flow in the reverse flow phase, when blood flows from the iliac arteries into the renal arteries. This altered path can be appreciated in Fig. \ref{fig:4B}, where the flow vectors at the caudal entrance of the renal artery are more cranially directed in the endograft models compared to control. In other words, renal entrance flow is less developed in the endograft models, leading to a more pronounced region of stagnant flow and adverse wall hemodynamics. With partial renal artery lumen coverage, graft material blocks incoming flow in the caudal renal artery, further distorting renal entrance flow, and allowing backflow near the caudal wall of the renal artery at small adverse pressure gradients. Graft fabric movement might alternatively effect flow dynamics. In the raw image data fabric movement of at most \SI{0.04}{mm} was present (data not shown), unlikely to affect overall renal entrance flow patterns. 
In the current setup, average renal flow of \SI{0.4}{L.min^{-1}} per side was unaltered despite the slight coverage from endograft material, as resistance at the renal and iliac outflow trajectories was adjusted to achieve equal flow distribution in all three models, to mimick physiologic arterial tone adjustment to achieve target-organ perfusion. In clinical practice, physiologic renal functional reserve\cite{Lao2011} may be suboptimal in EVAR patients, which could limit the extent to which renal autoregulation can compensate decreased renal inflow. Further impairment of renal flow by partial coverage may predispose to renal hypertension. Clinically established indicators of renal hypertension such as renal resistive index\citep{Mukherjee2001} are not applicable to the stent-induced coverage, as no upstream renal artery region was present in which such measures could have been quantified. 
An unexplained observation in the present study was the blunt flow profile that developed in the renal artery of the model with partial coverage. Interestingly, a similar flow velocity distribution with peak flow near the caudal renal wall has been observed in a Nellix stented model.\citep{Berger2000} The dissipation of momentum from the jet located cranially towards the caudal wall could not be explained by the current two-dimensional measurements. The flow was likely to exhibit regions with out-of-plane motion due to three-dimensional flow features such as vortices in the axial plane, which can impact WSS. The fast dissipation of momentum, along with the seemlingly irregular fluctuations in central lumen velocity (Fig. \ref{fig:5}), may also point to a localized region of turbulence, which in direct numerical simulations was shown to occur at low Reynolds numbers (\numrange{600}{1000}) near abrupt changes in geometry.\citep{Berger2000,Varghese2007}

\subsection{Limitations}
A reference for physiologic renal flow fields was not available, and temporal renal pressure data was not included in the current study. Considering the focus of the current study on renal flow, the temporal renal pressure profile was not investigated. The introduction of an invasive pressure catheter in the renal artery would have altered flow patterns therein. In addition, the pressure-flow relation has inherent simplifications due to the two-element or lower-order Windkessel models used in the current flow setup. Validating renal pressure profiles would therefore not have validated renal flow fields. Volumetric flows and pressures over time near the aortic bifurcation were collected in an earlier study performed with the same flow setup, showing a good match with reference data for flow and a reasonable match for pressure.\citep{GrootJebbink2014}
A second limitation of the present study is the lack of flow visualization outside the two-dimensional laser sheet. Besides a better understanding of the flow profile in the model with partial renal coverage, three-dimensional flow profiles could impact WSS at the proximal regions with stagnant flow. Still, with the reasonable assumption that turbulence is not present in this proximal region, the effects are likely to be small. PIV measurements in a plane orthogonal to the current, stereo-PIV\citep{Chen2014} and computational fluid dynamics could capture three-dimensional flow patterns with increasing detail. Furthermore, renal autoregulation could not be implemented in the measurements as its effect on renal blood flow is complex and multifactorial with both neurohumoral and tubuloglomerular feedback. In vivo PIV measurements with ultrasonography\citep{Faludi2010,Zhang2011} or 4D-flow MRI\citep{Markl2012} measurements could elucidate the effects of renal autoregulation on flow patterns, although both need substantial resolution improvements for doing so.

\section{Conclusion}
In current endovascular practice, unintended partial coverage of a renal artery may occur and leads to low and marked oscillations in WSS, potentially promoting atherosclerosis and subsequent renal artery stenosis. This may constitute a long-term clinical impact on renal function after EVAR. Awareness of the small offset often present between the edge of the graft fabric and their radiopaque markers is therefore imperative in endovascular practice. 

\section*{Acknowledgements}
We would like to thank the Radiology Department at the Rijnstate Hospital, Arnhem, the Netherlands for their contribution to this study by the implant sessions of the models. This work was supported in part by NanoNextNL, a micro and nanotechnology consortium of the Government of the Netherlands and 130 partners. In addition, we would like to thank Medtronic for providing the Endurant endografts. 

\bibliography{Revision}

\begin{thebibliography}{38}
\expandafter\ifx\csname natexlab\endcsname\relax\def\natexlab#1{#1}\fi
\providecommand{\url}[1]{\texttt{#1}}
\providecommand{\href}[2]{#2}
\providecommand{\path}[1]{#1}
\providecommand{\DOIprefix}{doi:}
\providecommand{\ArXivprefix}{arXiv:}
\providecommand{\URLprefix}{URL: }
\providecommand{\Pubmedprefix}{pmid:}
\providecommand{\doi}[1]{\href{http://dx.doi.org/#1}{\path{#1}}}
\providecommand{\Pubmed}[1]{\href{pmid:#1}{\path{#1}}}
\providecommand{\bibinfo}[2]{#2}
\ifx\xfnm\relax \def\xfnm[#1]{\unskip,\space#1}\fi
\bibitem[{Paravastu et~al.(2014)Paravastu, Jayarajasingam, Cottam, Palfreyman,
  Michaels, and Thomas}]{Paravastu2014}
\bibinfo{author}{S.~C.~V. Paravastu}, \bibinfo{author}{R.~Jayarajasingam},
  \bibinfo{author}{R.~Cottam}, \bibinfo{author}{S.~J. Palfreyman},
  \bibinfo{author}{J.~A. Michaels}, \bibinfo{author}{S.~M. Thomas},
\newblock \bibinfo{title}{{Endovascular repair of abdominal aortic aneurysm.}},
\newblock \bibinfo{journal}{The Cochrane database of systematic reviews}
  \bibinfo{volume}{1} (\bibinfo{year}{2014}) \bibinfo{pages}{CD004178}.
\bibitem[{Wald et~al.(2006)Wald, Waikar, Liangos, Pereira, Chertow, and
  Jaber}]{Wald2006}
\bibinfo{author}{R.~Wald}, \bibinfo{author}{S.~S. Waikar},
  \bibinfo{author}{O.~Liangos}, \bibinfo{author}{B.~J.~G. Pereira},
  \bibinfo{author}{G.~M. Chertow}, \bibinfo{author}{B.~L. Jaber},
\newblock \bibinfo{title}{{Acute renal failure after endovascular vs open
  repair of abdominal aortic aneurysm}},
\newblock \bibinfo{journal}{Journal of Vascular Surgery} \bibinfo{volume}{43}
  (\bibinfo{year}{2006}).
\bibitem[{Zarkowsky et~al.(2016)Zarkowsky, Hicks, Bostock, Stone, Eslami, and
  Goodney}]{Zarkowsky2016}
\bibinfo{author}{D.~S. Zarkowsky}, \bibinfo{author}{C.~W. Hicks},
  \bibinfo{author}{I.~C. Bostock}, \bibinfo{author}{D.~H. Stone},
  \bibinfo{author}{M.~Eslami}, \bibinfo{author}{P.~P. Goodney},
\newblock \bibinfo{title}{{Renal dysfunction and the associated decrease in
  survival after elective endovascular aneurysm repair}},
\newblock \bibinfo{journal}{Journal of Vascular Surgery} \bibinfo{volume}{64}
  (\bibinfo{year}{2016}) \bibinfo{pages}{1278--1285.e1}.
\bibitem[{Sun and Stevenson(2006)}]{Sun2006}
\bibinfo{author}{Z.~Sun}, \bibinfo{author}{G.~Stevenson},
\newblock \bibinfo{title}{{Transrenal Fixation of Aortic Stent-Grafts: Short-
  to Medterm Effects on Renal Function - A Systematic Review}},
\newblock \bibinfo{journal}{Radiology} \bibinfo{volume}{240}
  (\bibinfo{year}{2006}) \bibinfo{pages}{65--72}.
\bibitem[{Miller et~al.(2015)Miller, Razavi, and Lal}]{Miller2015}
\bibinfo{author}{L.~E. Miller}, \bibinfo{author}{M.~K. Razavi},
  \bibinfo{author}{B.~K. Lal},
\newblock \bibinfo{title}{{Suprarenal versus infrarenal stent graft fixation on
  renal complications after endovascular aneurysm repair}},
\newblock \bibinfo{journal}{Journal of Vascular Surgery} \bibinfo{volume}{61}
  (\bibinfo{year}{2015}) \bibinfo{pages}{1340--1349}.
\bibitem[{Walsh et~al.(2008)Walsh, Tang, and Boyle}]{Walsh2008a}
\bibinfo{author}{S.~R. Walsh}, \bibinfo{author}{T.~Y. Tang},
  \bibinfo{author}{J.~R. Boyle},
\newblock \bibinfo{title}{{Renal consequences of endovascular abdominal aortic
  aneurysm repair}},
\newblock \bibinfo{journal}{J Endovasc Ther} \bibinfo{volume}{15}
  (\bibinfo{year}{2008}) \bibinfo{pages}{73--82}.
\bibitem[{van Dijk et~al.(2003)van Dijk, van Sambeek, Cademartiri, and
  Pattynama}]{VanDijk2003}
\bibinfo{author}{L.~C. van Dijk}, \bibinfo{author}{M.~R. H.~M. van Sambeek},
  \bibinfo{author}{F.~Cademartiri}, \bibinfo{author}{P.~M.~T. Pattynama},
\newblock \bibinfo{title}{{Partial blockage of the renal artery ostium after
  stent-graft placement: detection and treatment.}},
\newblock \bibinfo{journal}{Journal of endovascular therapy : an official
  journal of the International Society of Endovascular Specialists}
  \bibinfo{volume}{10} (\bibinfo{year}{2003}) \bibinfo{pages}{684}.
\bibitem[{Schuurmann et~al.(2017)Schuurmann, Overeem, Ouriel, Slump, Jordan,
  Muhs, and {De Vries}}]{Schuurmann2017}
\bibinfo{author}{R.~C. Schuurmann}, \bibinfo{author}{S.~P. Overeem},
  \bibinfo{author}{K.~Ouriel}, \bibinfo{author}{C.~H. Slump},
  \bibinfo{author}{W.~D. Jordan}, \bibinfo{author}{B.~E. Muhs},
  \bibinfo{author}{J.~P.~P. {De Vries}},
\newblock \bibinfo{title}{{A Semiautomated Method for Measuring the
  3-Dimensional Fabric to Renal Artery Distances to Determine Endograft
  Position after Endovascular Aneurysm Repair}},
\newblock \bibinfo{journal}{Journal of Endovascular Therapy}
  (\bibinfo{year}{2017}).
\bibitem[{Ku et~al.(1985)Ku, Giddens, Zarins, and Glagov}]{Ku1985}
\bibinfo{author}{D.~N. Ku}, \bibinfo{author}{D.~P. Giddens},
  \bibinfo{author}{C.~K. Zarins}, \bibinfo{author}{S.~Glagov},
\newblock \bibinfo{title}{{Pulsatile Flow and Atherosclerosis in the Human
  Carotid Bifurcation Positive Correlation between Plaque Location and Low and
  Oscillating Shear Stress}},
\newblock \bibinfo{journal}{Arterioscler Thromb Vasc Biol} \bibinfo{volume}{5}
  (\bibinfo{year}{1985}) \bibinfo{pages}{293--302}.
\bibitem[{Moore et~al.(1994)Moore, Xu, Glagov, Zarins, and Ku}]{Moore1994b}
\bibinfo{author}{J.~E. Moore}, \bibinfo{author}{C.~Xu},
  \bibinfo{author}{S.~Glagov}, \bibinfo{author}{C.~K. Zarins},
  \bibinfo{author}{D.~N. Ku},
\newblock \bibinfo{title}{{Fluid wall shear stress measurements in a model of
  the human abdominal aorta: oscillatory behavior and relationship to
  atherosclerosis.}},
\newblock \bibinfo{journal}{Atherosclerosis} \bibinfo{volume}{110}
  (\bibinfo{year}{1994}) \bibinfo{pages}{225--240}.
\bibitem[{He and Ku(1996)}]{He1996}
\bibinfo{author}{X.~He}, \bibinfo{author}{D.~N. Ku},
\newblock \bibinfo{title}{{Pulsatile flow in the human left coronary artery
  bifurcation: average conditions.}},
\newblock \bibinfo{journal}{Journal of biomechanical engineering}
  \bibinfo{volume}{118} (\bibinfo{year}{1996}) \bibinfo{pages}{74--82}.
\bibitem[{Gimbrone and Garc{\'{i}}a-Carde{\~{n}}a(2013)}]{Gimbrone2013a}
\bibinfo{author}{M.~A. Gimbrone},
  \bibinfo{author}{G.~Garc{\'{i}}a-Carde{\~{n}}a},
\newblock \bibinfo{title}{{Vascular endothelium, hemodynamics, and the
  pathobiology of atherosclerosis}},
\newblock \bibinfo{journal}{Cardiovascular Pathology} \bibinfo{volume}{22}
  (\bibinfo{year}{2013}) \bibinfo{pages}{9--15}.
\bibitem[{Himburg et~al.(2004)Himburg, Grzybowski, Hazel, LaMack, Li, and
  Friedman}]{Himburg2004}
\bibinfo{author}{H.~A. Himburg}, \bibinfo{author}{D.~M. Grzybowski},
  \bibinfo{author}{A.~L. Hazel}, \bibinfo{author}{J.~A. LaMack},
  \bibinfo{author}{X.~M. Li}, \bibinfo{author}{M.~H. Friedman},
\newblock \bibinfo{title}{{Spatial comparison between wall shear stress
  measures and porcine arterial endothelial permeability}},
\newblock \bibinfo{journal}{Am J Physiol Heart Circ Physiol}
  \bibinfo{volume}{286} (\bibinfo{year}{2004}) \bibinfo{pages}{H1916--22}.
\bibitem[{Lee et~al.(2009)Lee, Antiga, and Steinman}]{Lee2009}
\bibinfo{author}{S.-W. Lee}, \bibinfo{author}{L.~Antiga},
  \bibinfo{author}{D.~A. Steinman},
\newblock \bibinfo{title}{{Correlations among indicators of disturbed flow at
  the normal carotid bifurcation.}},
\newblock \bibinfo{journal}{Journal of biomechanical engineering}
  \bibinfo{volume}{131} (\bibinfo{year}{2009}) \bibinfo{pages}{061013}.
\bibitem[{Bluestein et~al.(1996)Bluestein, Niu, Schoephoerster, and
  Dewanjee}]{Bluestein1996}
\bibinfo{author}{D.~Bluestein}, \bibinfo{author}{L.~Niu},
  \bibinfo{author}{R.~T. Schoephoerster}, \bibinfo{author}{M.~K. Dewanjee},
\newblock \bibinfo{title}{{Steady flow in an aneurysm model: correlation
  between fluid dynamics and blood platelet deposition}},
\newblock \bibinfo{journal}{Journal of Biomechanical Engineering}
  \bibinfo{volume}{118} (\bibinfo{year}{1996}) \bibinfo{pages}{280--286}.
\bibitem[{Katritsis et~al.(2010)Katritsis, Theodorakakos, Pantos, Andriotis,
  Efstathopoulos, Siontis, Karcanias, Redwood, and Gavaises}]{Katritsis2010}
\bibinfo{author}{D.~G. Katritsis}, \bibinfo{author}{A.~Theodorakakos},
  \bibinfo{author}{I.~Pantos}, \bibinfo{author}{A.~Andriotis},
  \bibinfo{author}{E.~P. Efstathopoulos}, \bibinfo{author}{G.~Siontis},
  \bibinfo{author}{N.~Karcanias}, \bibinfo{author}{S.~Redwood},
  \bibinfo{author}{M.~Gavaises},
\newblock \bibinfo{title}{{Vortex formation and recirculation zones in left
  anterior descending artery stenoses: computational fluid dynamics
  analysis.}},
\newblock \bibinfo{journal}{Physics in medicine and biology}
  \bibinfo{volume}{55} (\bibinfo{year}{2010}) \bibinfo{pages}{1395--411}.
\bibitem[{{Groot Jebbink} et~al.(2015){Groot Jebbink}, Grimme, Goverde, {Van
  Oostayen}, Slump, and Reijnen}]{GrootJebbink2015}
\bibinfo{author}{E.~{Groot Jebbink}}, \bibinfo{author}{F.~A.~B. Grimme},
  \bibinfo{author}{P.~C. J.~M. Goverde}, \bibinfo{author}{J.~A. {Van
  Oostayen}}, \bibinfo{author}{C.~H. Slump}, \bibinfo{author}{M.~M. P.~J.
  Reijnen},
\newblock \bibinfo{title}{{Geometrical consequences of kissing stents and the
  Covered Endovascular Reconstruction of the Aortic Bifurcation configuration
  in an in vitro model for endovascular reconstruction of aortic bifurcation}},
\newblock \bibinfo{journal}{Journal of Vascular Surgery} \bibinfo{volume}{61}
  (\bibinfo{year}{2015}) \bibinfo{pages}{1306--1311}.
\bibitem[{{Groot Jebbink} et~al.(2014){Groot Jebbink}, Goverde, van Oostayen,
  Reijnen, and Slump}]{GrootJebbink2014}
\bibinfo{author}{E.~{Groot Jebbink}}, \bibinfo{author}{P.~C. J.~M. Goverde},
  \bibinfo{author}{J.~A. van Oostayen}, \bibinfo{author}{M.~M. P.~J. Reijnen},
  \bibinfo{author}{C.~H. Slump},
\newblock \bibinfo{title}{{Innovation in aortoiliac stenting: an in vitro
  comparison}},
\newblock \bibinfo{journal}{Journal of Medical Imaging} \bibinfo{volume}{9036}
  (\bibinfo{year}{2014}) \bibinfo{pages}{90361X}.
\bibitem[{Moore and Ku(1994)}]{Moore1994}
\bibinfo{author}{J.~E. Moore}, \bibinfo{author}{D.~Ku},
\newblock \bibinfo{title}{{Pulsatile velocity measurements in a model of the
  human abdominal aorta under resting conditions}},
\newblock \bibinfo{journal}{Journal of Biomechanical Engineering}
  \bibinfo{volume}{116} (\bibinfo{year}{1994}) \bibinfo{pages}{337--346}.
\bibitem[{Yousif et~al.(2011)Yousif, Holdsworth, and Poepping}]{Yousif2011}
\bibinfo{author}{M.~Y. Yousif}, \bibinfo{author}{D.~W. Holdsworth},
  \bibinfo{author}{T.~L. Poepping},
\newblock \bibinfo{title}{{A blood-mimicking fluid for particle image
  velocimetry with silicone vascular models}},
\newblock \bibinfo{journal}{Experiments in Fluids} \bibinfo{volume}{50}
  (\bibinfo{year}{2011}) \bibinfo{pages}{769--774}.
\bibitem[{Scarano and Riethmuller(1999)}]{Scarano1999}
\bibinfo{author}{F.~Scarano}, \bibinfo{author}{M.~L. Riethmuller},
\newblock \bibinfo{title}{{Iterative multigrid approach in PIV image processing
  with discrete window offset}},
\newblock \bibinfo{journal}{Experiments in Fluids} \bibinfo{volume}{26}
  (\bibinfo{year}{1999}) \bibinfo{pages}{513--523}.
\bibitem[{Thielicke and Stamhuis(2014)}]{Thielicke2014}
\bibinfo{author}{W.~Thielicke}, \bibinfo{author}{E.~J. Stamhuis},
\newblock \bibinfo{title}{{PIVlab - Towards User-friendly, Affordable and
  Accurate Digital Particle Image Velocimetry in MATLAB}},
\newblock \bibinfo{journal}{Journal of Open Research Software}
  \bibinfo{volume}{2} (\bibinfo{year}{2014}) \bibinfo{pages}{e30}.
\bibitem[{Boersen et~al.(2017)Boersen, {Groot Jebbink}, Versluis, Slump, Ku,
  de~Vries, and Reijnen}]{Boersen2017b}
\bibinfo{author}{J.~T. Boersen}, \bibinfo{author}{E.~{Groot Jebbink}},
  \bibinfo{author}{M.~Versluis}, \bibinfo{author}{C.~H. Slump},
  \bibinfo{author}{D.~N. Ku}, \bibinfo{author}{J.-P.~P. de~Vries},
  \bibinfo{author}{M.~M. P.~J. Reijnen},
\newblock \bibinfo{title}{{Flow and wall shear stress characterization
  following endovascular aneurysm repair and endovascular aneurysm sealing in
  an infrarenal aneurysm model}},
\newblock \bibinfo{journal}{Journal of Vascular Surgery} \bibinfo{volume}{In
  press} (\bibinfo{year}{2017}).
\bibitem[{Peiffer et~al.(2013)Peiffer, Sherwin, and Weinberg}]{Peiffer2013a}
\bibinfo{author}{V.~Peiffer}, \bibinfo{author}{S.~J. Sherwin},
  \bibinfo{author}{P.~D. Weinberg},
\newblock \bibinfo{title}{{Does low and oscillatory wall shear stress correlate
  spatially with early atherosclerosis? A systematic review}},
\newblock \bibinfo{journal}{Cardiovascular Research} \bibinfo{volume}{99}
  (\bibinfo{year}{2013}) \bibinfo{pages}{242--250}.
\bibitem[{Morbiducci et~al.(2010)Morbiducci, Gallo, Massai, Consolo, Ponzini,
  Antiga, Bignardi, Deriu, and Redaelli}]{Morbiducci2010}
\bibinfo{author}{U.~Morbiducci}, \bibinfo{author}{D.~Gallo},
  \bibinfo{author}{D.~Massai}, \bibinfo{author}{F.~Consolo},
  \bibinfo{author}{R.~Ponzini}, \bibinfo{author}{L.~Antiga},
  \bibinfo{author}{C.~Bignardi}, \bibinfo{author}{M.~a. Deriu},
  \bibinfo{author}{A.~Redaelli},
\newblock \bibinfo{title}{{Outflow conditions for image-based hemodynamic
  models of the carotid bifurcation: implications for indicators of abnormal
  flow.}},
\newblock \bibinfo{journal}{Journal of biomechanical engineering}
  \bibinfo{volume}{132} (\bibinfo{year}{2010}) \bibinfo{pages}{091005}.
\bibitem[{Suess et~al.(2016)Suess, Anderson, Danielson, Pohlson, Remund,
  Blears, Gent, and Kelly}]{Suess2016}
\bibinfo{author}{T.~Suess}, \bibinfo{author}{J.~Anderson},
  \bibinfo{author}{L.~Danielson}, \bibinfo{author}{K.~Pohlson},
  \bibinfo{author}{T.~Remund}, \bibinfo{author}{E.~Blears},
  \bibinfo{author}{S.~Gent}, \bibinfo{author}{P.~Kelly},
\newblock \bibinfo{title}{{Examination of near-wall hemodynamic parameters in
  the renal bridging stent of various stent graft configurations for repairing
  visceral branched aortic aneurysms}},
\newblock \bibinfo{journal}{Journal of Vascular Surgery} \bibinfo{volume}{64}
  (\bibinfo{year}{2016}) \bibinfo{pages}{788--796}.
\bibitem[{Casa et~al.(2015)Casa, Deaton, and Ku}]{Casa2015}
\bibinfo{author}{L.~D.~C. Casa}, \bibinfo{author}{D.~H. Deaton},
  \bibinfo{author}{D.~N. Ku},
\newblock \bibinfo{title}{{Role of high shear rate in thrombosis}},
\newblock \bibinfo{journal}{Journal of Vascular Surgery} \bibinfo{volume}{61}
  (\bibinfo{year}{2015}) \bibinfo{pages}{1068--1080}.
\bibitem[{Eshtehardi et~al.(2017)Eshtehardi, Brown, Bhargava, Costopoulos,
  Hung, Corban, Hosseini, Gogas, Giddens, and Samady}]{Eshtehardi2017}
\bibinfo{author}{P.~Eshtehardi}, \bibinfo{author}{A.~J. Brown},
  \bibinfo{author}{A.~Bhargava}, \bibinfo{author}{C.~Costopoulos},
  \bibinfo{author}{O.~Y. Hung}, \bibinfo{author}{M.~T. Corban},
  \bibinfo{author}{H.~Hosseini}, \bibinfo{author}{B.~D. Gogas},
  \bibinfo{author}{D.~P. Giddens}, \bibinfo{author}{H.~Samady},
\newblock \bibinfo{title}{{High wall shear stress and high-risk plaque: an
  emerging concept}},
\newblock \bibinfo{journal}{International Journal of Cardiovascular Imaging}
  \bibinfo{volume}{33} (\bibinfo{year}{2017}) \bibinfo{pages}{1089--1099}.
\bibitem[{Holenstein and Ku(1988)}]{Holenstein1988}
\bibinfo{author}{R.~Holenstein}, \bibinfo{author}{D.~N. Ku},
\newblock \bibinfo{title}{{Reverse flow in the major infrarenal vessels--a
  capacitive phenomenon.}},
\newblock \bibinfo{journal}{Biorheology} \bibinfo{volume}{25}
  (\bibinfo{year}{1988}) \bibinfo{pages}{835--42}.
\bibitem[{Walsh et~al.(2003)Walsh, Chin-Quee, and Moore}]{Walsh2003}
\bibinfo{author}{P.~W. Walsh}, \bibinfo{author}{S.~Chin-Quee},
  \bibinfo{author}{J.~E. Moore},
\newblock \bibinfo{title}{{Flow changes in the aorta associated with the
  deployment of a AAA stent graft}},
\newblock \bibinfo{journal}{Medical Engineering and Physics}
  \bibinfo{volume}{25} (\bibinfo{year}{2003}) \bibinfo{pages}{299--307}.
\bibitem[{Lao et~al.(2011)Lao, Parasher, Cho, and Yeghiazarians}]{Lao2011}
\bibinfo{author}{D.~Lao}, \bibinfo{author}{P.~S. Parasher},
  \bibinfo{author}{K.~C. Cho}, \bibinfo{author}{Y.~Yeghiazarians},
\newblock \bibinfo{title}{{Atherosclerotic renal artery stenosis - Diagnosis
  and treatment}},
\newblock \bibinfo{journal}{Mayo Clinic Proceedings} \bibinfo{volume}{86}
  (\bibinfo{year}{2011}) \bibinfo{pages}{649--657}.
\bibitem[{Mukherjee et~al.(2001)Mukherjee, Bhatt, Robbins, Roffi, Cho,
  Reginelli, Bajzer, Navarro, and Yadav}]{Mukherjee2001}
\bibinfo{author}{D.~Mukherjee}, \bibinfo{author}{D.~L. Bhatt},
  \bibinfo{author}{M.~Robbins}, \bibinfo{author}{M.~Roffi},
  \bibinfo{author}{L.~Cho}, \bibinfo{author}{J.~Reginelli},
  \bibinfo{author}{C.~Bajzer}, \bibinfo{author}{F.~Navarro},
  \bibinfo{author}{J.~S. Yadav},
\newblock \bibinfo{title}{{Renal artery end-diastolic velocity and renal artery
  resistance index as predictors of outcome after renal stenting}},
\newblock \bibinfo{journal}{American Journal of Cardiology}
  \bibinfo{volume}{88} (\bibinfo{year}{2001}) \bibinfo{pages}{1064--1066}.
\bibitem[{Berger and Jou(2000)}]{Berger2000}
\bibinfo{author}{S.~A. Berger}, \bibinfo{author}{L.-d. Jou},
\newblock \bibinfo{title}{{Flows in Stenotic Vessels}},
\newblock \bibinfo{journal}{Annual Review of Fluid Mechanics}
  \bibinfo{volume}{32} (\bibinfo{year}{2000}) \bibinfo{pages}{347--382}.
\bibitem[{Varghese et~al.(2007)Varghese, Frankel, and Fischer}]{Varghese2007}
\bibinfo{author}{S.~S. Varghese}, \bibinfo{author}{S.~H. Frankel},
  \bibinfo{author}{P.~F. Fischer},
\newblock \bibinfo{title}{{Direct numerical simulation of stenotic flows, Part
  2: Pulsatile flow}},
\newblock \bibinfo{journal}{Journal of Fluid Mechanics} \bibinfo{volume}{582}
  (\bibinfo{year}{2007}) \bibinfo{pages}{281}.
\bibitem[{Chen et~al.(2014)Chen, Ant{\'{o}}n, Hung, Menon, Finol, and
  Pekkan}]{Chen2014}
\bibinfo{author}{C.-Y. Chen}, \bibinfo{author}{R.~Ant{\'{o}}n},
  \bibinfo{author}{M.-y. Hung}, \bibinfo{author}{P.~Menon},
  \bibinfo{author}{E.~a. Finol}, \bibinfo{author}{K.~Pekkan},
\newblock \bibinfo{title}{{Effects of intraluminal thrombus on patient-specific
  abdominal aortic aneurysm hemodynamics via stereoscopic particle image
  velocity and computational fluid dynamics modeling.}},
\newblock \bibinfo{journal}{Journal of biomechanical engineering}
  \bibinfo{volume}{136} (\bibinfo{year}{2014}) \bibinfo{pages}{031001}.
\bibitem[{Faludi et~al.(2010)Faludi, Szulik, D'hooge, Herijgers, Rademakers,
  Pedrizzetti, and Voigt}]{Faludi2010}
\bibinfo{author}{R.~Faludi}, \bibinfo{author}{M.~Szulik},
  \bibinfo{author}{J.~D'hooge}, \bibinfo{author}{P.~Herijgers},
  \bibinfo{author}{F.~Rademakers}, \bibinfo{author}{G.~Pedrizzetti},
  \bibinfo{author}{J.~U. Voigt},
\newblock \bibinfo{title}{{Left ventricular flow patterns in healthy subjects
  and patients with prosthetic mitral valves: An in vivo study using
  echocardiographic particle image velocimetry}},
\newblock \bibinfo{journal}{Journal of Thoracic and Cardiovascular Surgery}
  \bibinfo{volume}{139} (\bibinfo{year}{2010}) \bibinfo{pages}{1501--1510}.
\bibitem[{Zhang et~al.(2011)Zhang, Lanning, Mazzaro, Barker, Gates, Strain,
  Fulford, Gosling, Shore, Bellenger, Rech, Chen, Chen, and
  Shandas}]{Zhang2011}
\bibinfo{author}{F.~Zhang}, \bibinfo{author}{C.~Lanning},
  \bibinfo{author}{L.~Mazzaro}, \bibinfo{author}{A.~J. Barker},
  \bibinfo{author}{P.~E. Gates}, \bibinfo{author}{W.~D. Strain},
  \bibinfo{author}{J.~Fulford}, \bibinfo{author}{O.~E. Gosling},
  \bibinfo{author}{A.~C. Shore}, \bibinfo{author}{N.~G. Bellenger},
  \bibinfo{author}{B.~Rech}, \bibinfo{author}{J.~Chen},
  \bibinfo{author}{J.~Chen}, \bibinfo{author}{R.~Shandas},
\newblock \bibinfo{title}{{In vitro and preliminary in vivo validation of echo
  particle image velocimetry in carotid vascular imaging}},
\newblock \bibinfo{journal}{Ultrasound in Medicine and Biology}
  \bibinfo{volume}{37} (\bibinfo{year}{2011}) \bibinfo{pages}{450--464}.
\bibitem[{Markl et~al.(2012)Markl, Frydrychowicz, Kozerke, Hope, and
  Wieben}]{Markl2012}
\bibinfo{author}{M.~Markl}, \bibinfo{author}{A.~Frydrychowicz},
  \bibinfo{author}{S.~Kozerke}, \bibinfo{author}{M.~Hope},
  \bibinfo{author}{O.~Wieben},
\newblock \bibinfo{title}{{4D flow MRI}},
\newblock \bibinfo{journal}{Journal of Magnetic Resonance Imaging}
  \bibinfo{volume}{36} (\bibinfo{year}{2012}) \bibinfo{pages}{1015--1036}.

\end{thebibliography}
\end{document}